\documentclass[letterpaper]{article} 
\usepackage{aaai2026}  
\usepackage{times}  
\usepackage{helvet}  
\usepackage{courier}  
\usepackage[hyphens]{url}  
\usepackage{graphicx} 
\usepackage{natbib}  
\usepackage{caption} 
\usepackage[utf8]{inputenc}
\makeatletter
\def\copyright@on{F}
\makeatother
\def\showauthors@on{T}

\usepackage{algorithm}
\usepackage{algorithmic}
\usepackage{amssymb} 

\usepackage{amsmath}
\usepackage{multirow}
\usepackage{booktabs} 

\usepackage{newfloat}
\usepackage{listings}
\DeclareCaptionStyle{ruled}{labelfont=normalfont,labelsep=colon,strut=off} 
\floatstyle{ruled}
\newfloat{listing}{tb}{lst}{}
\floatname{listing}{Listing}
\title{FlowCrypt: Flow-Based Lightweight Encryption with Near-Lossless Recovery for Cloud Photo Privacy}
\author {
    Xiaohui Yang,
    Ping Ping,
    Feng Xu
}
\affiliations {
    Hohai University
}

\begin{document}

\maketitle

\begin{abstract}
The widespread adoption of smartphone photography has led users to increasingly rely on cloud storage for personal photo archiving and sharing, raising critical privacy concerns. Existing deep learning-based image encryption schemes, typically built upon CNNs or GANs, often depend on traditional cryptographic algorithms and lack inherent architectural  reversibility, resulting in limited recovery quality and poor robustness. Invertible neural networks (INNs) have emerged to address this issue by enabling reversible transformations, yet the first INN-based encryption scheme still relies on an auxiliary reference image and discards by-product information before decryption, leading to degraded recovery and limited practicality. To address these limitations, this paper proposes FlowCrypt, a novel flow-based image encryption framework that simultaneously achieves near-lossless recovery, high security, and lightweight model design. FlowCrypt begins by applying a key-conditioned random split to the input image, enhancing forward-process randomness and encryption strength. The resulting components are processed through a Flow-based Encryption/Decryption (FED) module composed of invertible blocks, which share parameters across encryption and decryption. Thanks to its reversible architecture and reference-free design, FlowCrypt ensures high-fidelity image recovery. Extensive experiments show that FlowCrypt achieves recovery quality with 100dB on three datasets, produces uniformly distributed cipher images, and maintains a compact architecture with only 1M parameters, making it suitable for mobile and edge-device applications.
\end{abstract}


\section{Introduction}

With the advancement of built-in smartphone cameras, the number of high-quality personal photos has grown dramatically. Due to the limited local storage capacity and the risk of accidental data loss, many users consider to rely on cloud storage services provided by companies such as Google, Apple iCloud, and others for long-term storage and sharing of private photos. While cloud storage greatly facilitates convenient photo backup and sharing, it also raises increasing concerns about potential privacy breaches, making the security of cloud photo storage an urgent issue. 

Image encryption offers an effective solution to protect sensitive photo content from unauthorized access during cloud storage and transmission. By converting a meaningful plain image into an unintelligible, noise-like cipher image, encryption ensures that only authorized users with the correct key can recover the original image. Early encryption algorithms such as Data Encryption Standard (DES)  \cite{1970DES}, International Data Encryption Algorithm (IDEA) \cite{Lai1991IDEA} and Advanced Encryption Standard (AES) \cite{2001AES}, were originally designed for general-purpose digital data including text and other binary forms. While these ciphers remain secure and efficient with modern hardware acceleration, they operate at the byte or block level without exploiting the spatial redundancy and statistical characteristics inherent in images, which limits their suitability in scenarios requiring both strong security and image-specific objectives. To address these limitations, various specialized image encryption schemes have been proposed, including methods based on chaotic system \cite{HUA20152DSLMM, HUA20162DLASM}, cellular automata (CA), DNA encoding, wave transform, optical encryption and elliptic curves, which offer desirable properties such as non-linearity, sensitivity to initial conditions, and high efficiency tailored to image data. More recently, advances in deep learning have inspired the utilization of deep neural networks (DNNs) for image encryption, leveraging their powerful nonlinear transformations and semantic extraction capabilities. Some schemes utilize DNNs to generate encryption keys \cite{Erkan_2022_CNNRIE, 2023_GANRIE}, which are then combined with traditional algorithms, while others employ DNNs to produce an initial cipher image that is further refined by conventional techniques \cite{2017CycleGAN, 2021DingDeepEDN}. Despite these advances, most DNN-based methods still resort to traditional cryptography, limiting their potential as fully end-to-end encryption solutions and posing challenges for robustness and security.

\begin{figure}[t]
\centering
\includegraphics[width=1.0\columnwidth]{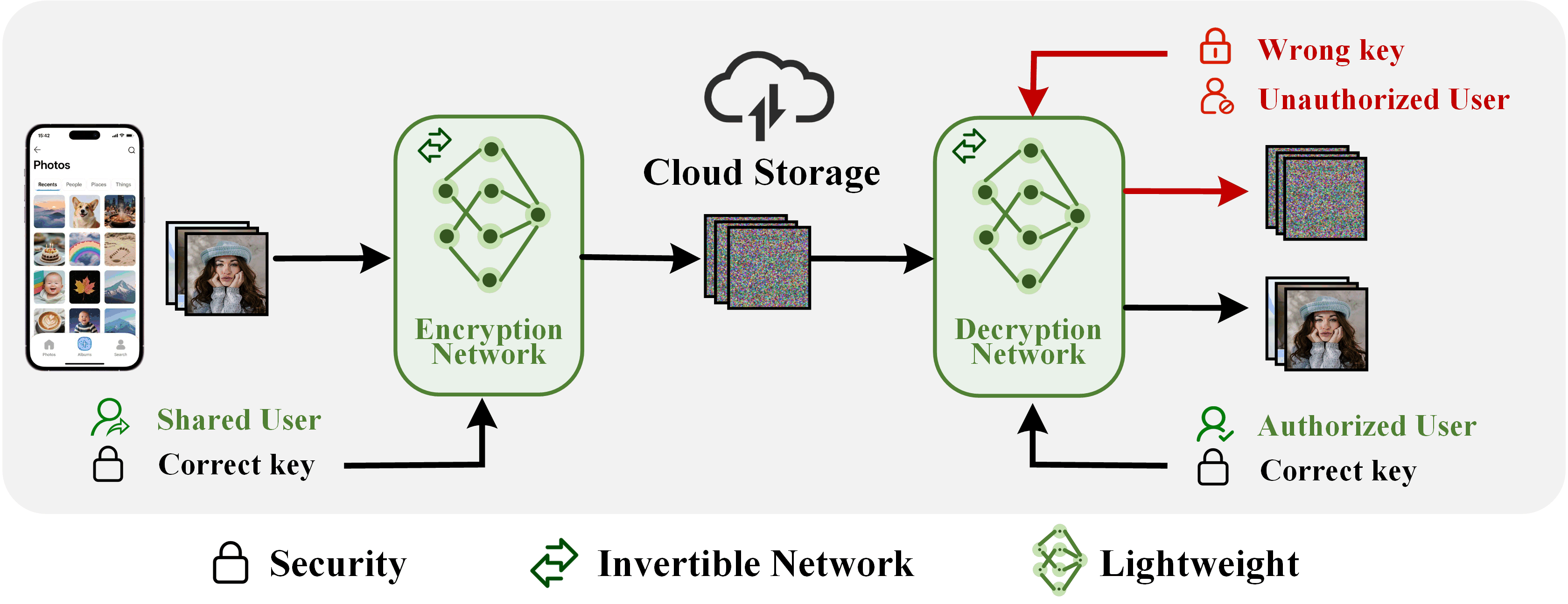} 
\caption{A paradigm of FlowCrypt applied in cloud photo privacy.}
\label{fig1}
\end{figure}

To address these limitations, invertible neural networks (INNs) have emerged as a promising direction for reversible image processing tasks, such as image encryption \cite{INN_RIE_2025}, image hiding \cite{2021HiNet}, and digital watermarking \cite{2023FlowRobustWatermarking}. Recent work has proposed INN-based framework \cite{INN_RIE_2025} for complete end-to-end image encryption using shared encryptor and decryptor. Nevertheless, this INN-based method has yet to achieve truly lossless decryption, as it typically rely on an additional reference image, which is a pre-selected random noise image, and discard by-product information before decryption, which limits recovery fidelity and practicality.

To achieve high recovery fidelity and secure cloud photo encryption, we propose \textbf{FlowCrypt}, 
a novel \textbf{flow-based image encryption framework} that simultaneously ensures \textbf{near lossless recovery}, \textbf{high security}, and \textbf{lightweight} design. As illustrated in Fig. \ref{fig1}, a user who intends to share private photos via cloud storage first encrypts the image using an invertible neural network with approximately 1M parameters. The resulting cipher image appears noise-like and semantically meaningless, effectively protecting sensitive content during cloud transmission and storage. Thanks to the intrinsic reversibility of the flow-based model, the original image can be near losslessly recovered using the same model parameters. Furthermore, encryption is conditioned on a carefully designed secret-key mechanism, ensuring that only authorized users with the correct key can decrypt the image. Unlike prior INN-based encryption scheme, FlowCrypt does not require any auxiliary reference image and not discard any information, improving recovery fidelity and practical deployability. FlowCrypt strikes a favorable balance between high-fidelity recovery, robust security, and lightweight model payload, making it highly suitable for privacy protection in resource-constrained environments such as smartphones and edge devices. The key contributions of this paper are summarized as follows:

\begin{itemize}
    \item \textbf{Near-lossless recovery.} Leveraging the inherent invertibility of flow-based model and preserving all transformation information during encryption, FlowCrypt achieves near-lossless image recovery without relying on auxiliary reference images.
    
    \item \textbf{High security.} The reference-free cipher images exhibit uniform distribution and high entropy, ensuring robustness against statistical attacks. The specially crafted secret-key mechanism guarantees that only authorized users with the correct key can successfully decrypt the image.
      
    \item \textbf{Lightweight design.} The whole framework is composed of a compact network structure with approximately 1 million parameters, significantly reducing computational overhead. This lightweight design enables real-time performance and facilitates deployment on mobile devices and other resource-constrained platforms
\end{itemize}

\section{Related Work}

\subsection{Image Encryption}

Image encryption aims to protect visual content by transforming plain images into noise-like, unintelligible forms through pixel distortion. Most traditional encryption methods were inspired by \cite{1997_IE_chatoticmap}, which first introduced the use of chaotic maps such as logistic map to drive image encryption. Since then, various one-dimensional (1D) chaotic maps (e.g., logistic, sine, and tent maps) have been explored. Although multi-dimensional chaotic maps offer more complex structures and chaotic behaviors, they often incur higher computational costs. As a compromise between complexity and efficiency, two-dimensional (2D) chaotic maps have gained popularity. Representative workings include the 2D Logistic-Sine-Coupling Map \cite{HUA20182DLogistic}, 2D Sine Logistic Modulation Map \cite{HUA20152DSLMM}, 2D Logistic-Adjusted-Sine Map \cite{HUA20162DLASM}, and the 2D Logistic Tent Modular Map for color image encryption \cite{HUA20212DLogisticTent}.

In recent years, deep neural network (DNN)-based encryption schemes have emerged. 
These methods exploit the non-linearity and semantic extraction capabilities of DNNs to either generate secret keys or directly perform encryption. In key generation schemes, DNNs are used solely to produce secret keys, while encryption follows traditional chaotic-based procedures. Erkan et al. \cite{Erkan_2022_CNNRIE} first combined convolutional neural networks (CNNs) with chaotic maps for encryption. Singh \cite{2023_GANRIE} further enhanced this idea by using a generative adversarial network (GAN) to generate keys, followed by scrambling and diffusion operations. However, these approaches treat DNNs merely as auxiliary components, the encryption effect is still achieved by traditional schemes. To leverage DNNs more directly, researchers have begun reframing image encryption as a task of transferring an image into a visually inconsistent one \cite{DeepEDN_2021}. For example, CycleGAN \cite{2017CycleGAN} has been employed for medical image encryption, while CNN-based methods \cite{2022FEDResNet} perform initial encryption, with chaotic systems subsequently enhancing security. Nonetheless, the reliance on chaotic systems means these DNN-based schemes cannot be regarded as complete end-to-end DNN-based encryption methods and often suffer from poor robustness. To address these limitations, an invertible neural network (INN)-based encryption scheme \cite{INN_RIE_2025} is proposed for complete end-to-end robust image encryption, employing a shared encryptor and decryptor to enable reversible transformation. While this scheme represents meaningful progress, it still depends on a reference random-noise image as additional guidance and fails to achieve ideal recovery quality.

\subsection{Normalization Flow-based Model}

Normalizing flow-based model can directly compute the likelihoods and is widely applied in generative tasks. Flow-based model is built upon invertible neural networks, which allows both forward and backward mappings using the same network and parameters. Leveraging the reversibility of INN, several models have demonstrated powerful generative capabilities, including NICE \cite{dinh2015nice}, RealNVP \cite{dinh2017RealNVP}, Glow \cite{2018Glow}, and i-RevNet \cite{2018irevnet}. The inherent reversibility of flow-based model makes it particularly well-suited for image hiding \cite{2021HiNet}, robust watermarking \cite{2023FlowRobustWatermarking}, and image encryption \cite{INN_RIE_2025} tasks, as it can effectively achieve the perfect recovery.

\section{Proposed Framework}

\subsection{Overview}

\begin{figure*}[t]
\centering
\includegraphics[width=1.0\textwidth]{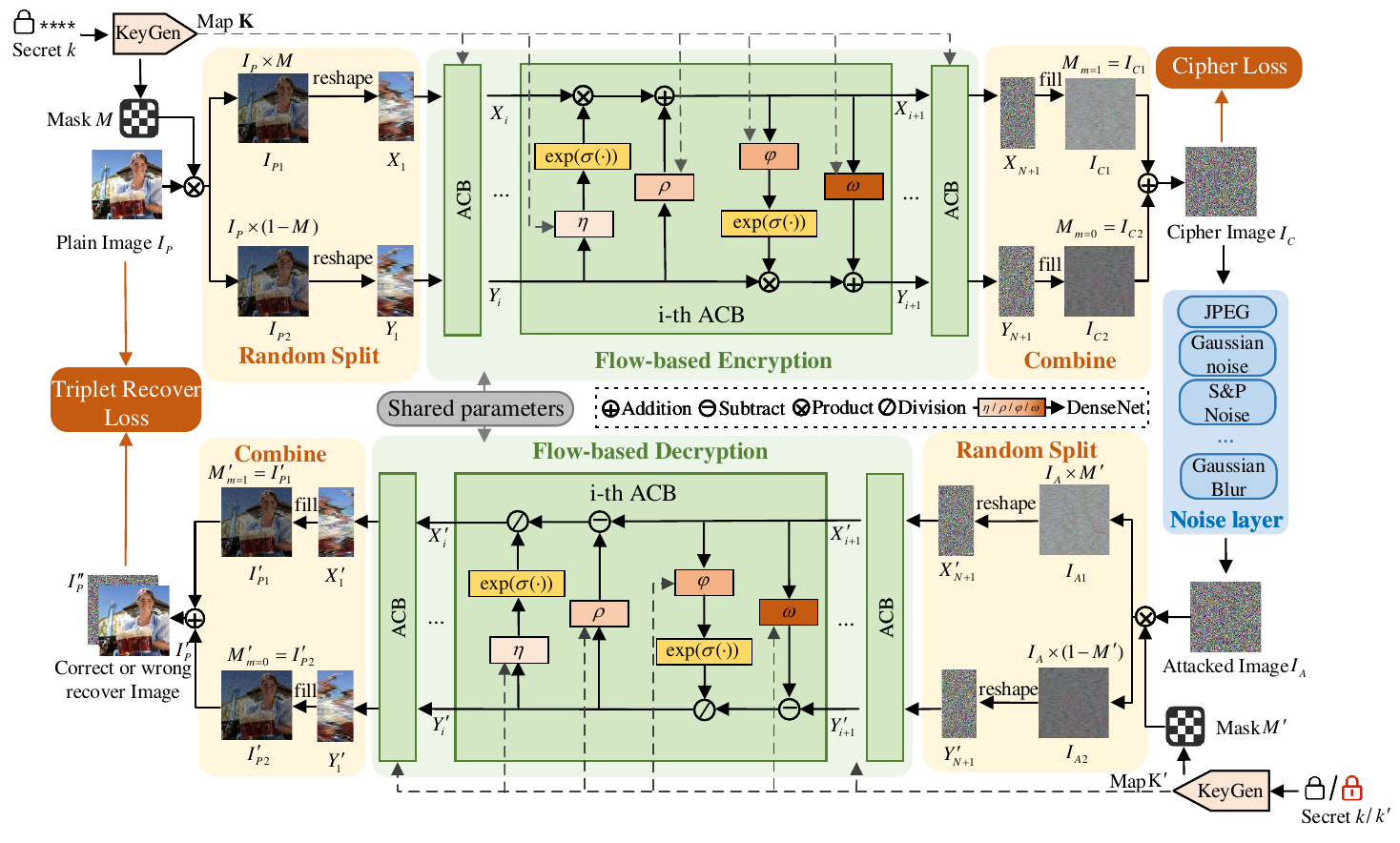} 
\caption{The framework of FlowCrypt comprises flow-based encryption and decryption, built upon three core components: key generation and random split strategy, affine coupling blocks (ACBs), and noise layer.}
\label{fig2}
\end{figure*}

The overall workflow of FlowCrypt is illustrated in Fig.\ref{fig2}. This framework aims to construct an invertible image encryption framework that simultaneously achieves high security, nearly lossless recovery, and lightweight model. Given a plain image $I_P$, a random split strategy conditioned on secret keys is first applied to divide it into two components $X_{1}$ and $X_{2}$. This step injects randomness and improves the unpredictability of the forward encryption process, thereby enhancing security. The resulting components are then passed through the flow-based encryption/decryption (FED) module, which is composed of several invertible neural blocks built upon the affine coupling block (ACB) structure, well-suited for reversible image encryption and decryption tasks. These blocks further incorporate key-conditioned randomness to strengthen encryption security. The FED module transforms $X_1$ and $X_2$ into encrypted representations $X_{N+1},Y_{N+1}$, which are then combined to generate the cipher image $I_C$. To enhance robustness under real-world conditions, a noise simulation layer is introduced to distort the cipher image $I_C$, producing an degraded image $I_A$ for robust decryption training. The backward decryption takes $I_A$ as input and reconstructs the recovery $I^{'}_P$ via the inverse flow of the FED module. Due to the reversible architecture of FED and less discarded information enabled by the no-reference encryption mechanism, near-lossless recovery can be guaranteed. Furthermore, the FED module shares parameters for both forward encryption and backward decryption, contributing to the model's lightweight design. The detailed design of each key component within FlowCrypt is elaborated as below.

\subsection{Key Generation and Random Split Strategy}

The key generation module (KeyGen) is designed to produce pseudorandom, key-independent information that introduce randomness into the FED module. Given a user-defined secret key $k$, which serves as a password with arbitrary length and characters, a Password-Based Key Derivation Function (PBKDF) is applied to generate a reproducible pseudorandom bitstream with strong key sensitivity. In particular, we adopt PBKDF2 with a high iteration count and a fixed salt to derive a byte stream of length $W \times H$, where $W$ and $H$ are the width and height of the input image $I_P$. 
To ensure the generated binary mask is both key-dependent and strictly balanced (i.e., with equal numbers of 0s and 1s), we apply a two-step strategy (see Appendix A.1 for details). First, we generate a balanced binary vector consisting of exactly half 0s and half 1s. Then, the pseudorandom byte stream derived from PBKDF is used to deterministically shuffle this vector, resulting in a 2D binary mask $M\in \{0,1\}^{W\times H}$, which is used to inject key-independent randomness into the following encryption process:
\begin{equation}
    M=\mathcal{T}(\text{PBKDF}(k)), \quad M\in \{0,1\}^{W\times H}
\end{equation}
where $\mathcal{T}(.)$ denotes the binarization and balancing transformation applied to the PBKDF output.

The resulting binary mask $M$ is then applied to match each channel of the input image $I_O$ and spatially partition $I_O$ into two complementary parts $I_1$ and $I_2$ of the same size:
\begin{equation}
    I_{P1}=M\times I_P, \quad I_{P2}=(1-M)\times I_P
\end{equation}
where $I_{P1}$ and $I_{P2}$ retain mutually exclusive pixel subsets of $I_P$, while maintaining the same spatial dimensions.

To prepare the data for invertible encryption, the non-zero pixels in $I_{P1}$ and $I_{P2}$ are reorganized into two tensors $X_1$ and $Y_1$ with dimensions $(3,H,W/2)$, using a row-wise and column-wise ordering fashion, respectively. These serve as the paired inputs to the subsequent flow-based encryption module (FED). After the FED processes the inputs and produces the encrypted representations $X_{N+1}$ and $Y_{N+1}$, the original spatial structure is restored based on a binary mask $M$. Specifically, a zero-initialized tensor of shape ($H\times W$) is used to merge the encrypted components: positions corresponding to mask value 1 are filled with $X_{N+1}$, and those corresponding to 0 are filled with $Y_{N+1}$. This strategy ensures that the generated mask is deterministically reproducible from the user-specific password, strictly balanced to preserve the image information symmetry, and highly sensitive to even minor changes in the initial secret key due to the cryptographic properties of PBKDF. Consequently, it enhances the security and unpredictability of the FED module, effectively preventing reverse-engineering of the key or mask from the cipher data. 

\subsection{Flow-based Encryption/Decryption}

 Thanks to the strict mathematical reversibility, flow-based network is inherently suitable for image encryption and decryption tasks. The proposed flow-based encryption/decryption module (FED) consists of two primary components: the forward encryption function $f_\theta$ and its inverse $f_\theta^{-1}$, which represents backward decryption. Both functions share the same set of parameters $\theta$. In the forward encryption process, two tensors $X_1,Y_1 \in \mathbb{R}^{3\times H \times \frac{W}2}$ derived from random split module, are sequentially processed by $N$ invertible neural blocks (INBs), yielding two transformed tensors $X_{N+1},Y_{N+1} \in \mathbb{R}^{3\times H \times \frac{W}2}$, which are then fused to generate the final cipher image. For backward decryption, two input tensors  $X^{'}_{N+1},Y^{'}_{N+1} \in \mathbb{R}^{3\times H \times \frac{W}2}$ extracted from the degraded image $I_\mathcal{A}$ are passed through the same sequence of INBs in reverse order. This yields the tensors $X^{'}_1,Y^{'}_1 \in \mathbb{R}^{3\times H \times \frac{W}2}$ which can further combined to reconstruct the recovered image $I^{'}_P$.
 
 INBs are the core components of the FED framework. As illustrated in Fig.\ref{fig2}, each INB comprises four non-linear mapping functions $\eta()$, $\rho()$, $\varphi()$ and $\omega()$. Each function adopts the same densely connected Conv-LeakyReLu structure, but with independent parameters. Theses functions collectively form a bijective affine coupling block (ACB), ensuring invertibility. Importantly, each mapping function receives not only the feature maps from plain image, but also secret map $\mathbf{K}$ generated from a secret key $k$. The secret map $\mathbf{K}$ is concatenated with the image feature maps $X_i,Y_i$ in a channel-wise manner (e.g. replace $\varphi(Y_i)$ with $\varphi (Y_i || \mathbf{K})$ etc). Each nonlinear mapping function maintains the spatial dimensions of the input image feature maps:
 \begin{equation}
    \eta,\rho,\varphi,\omega: \mathbb{R}^{(C+1)\times H \times \frac{W}2} \to \mathbb{R}^{C\times H \times \frac{W}2}
 \end{equation}
 where $C=3$, representing the number of RGB image channels. The transformation within the $i^{th}$ INB during forward encryption process is formulated as:
 \begin{equation}
 \begin{split}
    X_{i+1}&=X_i \cdot \text{exp}(\sigma(\eta(Y_i||\mathbf{K}))) + \rho(Y_i||\mathbf{K}) \\
    Y_{i+1}&=Y_i \cdot \text{exp}(\sigma(\varphi(X_{i+1}||\mathbf{K}))) + \omega(X_{i+1}||\mathbf{K})
 \end{split}
 \end{equation}
 where $i \in [1 \dots N]$, $||$ denotes channel-wise concatenation and $\sigma (\cdot)$ denotes the sigmoid activation, which served to constrain the image feature map. After being processed by $N$ invertible neural blocks as such, we can obtain $X_{N+1}$ and $Y_{N+1}$ which used to combine the final cipher image.

For backward decryption, the computation proceeds inversely through the INBs. In particular, for the first INB, the input data is the split result of the degraded image $I_{\mathcal{A}}$ distorted by the noise layer. For the $i^{th}$ INB, the recovery process between inputs $X^{'}_{i+1}, Y^{'}_{i+1}$ and outputs $X^{'}_{i}, Y^{'}_{i}$ is given by:
 \begin{equation}
 \begin{split}
    Y^{'}_{i}&=(Y^{'}_{i+1} - \omega(X^{'}_{i+1}||\mathbf{K}^{'})) \cdot \text{exp}(-\sigma(\varphi(X^{'}_{i+1}||\mathbf{K}^{'}))) \\
    X^{'}_{i}&=(X^{'}_{i+1} - \rho(Y^{'}_i||\mathbf{K}^{'})) \cdot \text{exp}(-\sigma(\eta(Y^{'}_i||\mathbf{K}^{'})))
 \end{split}
 \end{equation}
 where secret map $\mathbf{K}^{'}$ is regenerated identically using the same secret key $k$. After passing through all $N$ invertible neural blocks, the final outputs $X^{'}_1$ and $Y^{'}_1$ are merged to reconstruct the recovered image. To ensure security, the model is trained such that correct decryption is only possible with the matching key $k$. When an incorrect key $k^{'}$ is used, the recovered image becomes significantly distorted and visually unrecognizable. In this paper, each non-linear mapping function comprises a stack of 5 "Conv-LeakyReLu" blocks, as illustrated in Fig.\ref{fig2}.
 
\subsection{Noise Layer}

In daily image transmission, the cipher image $I_C$ is often degraded by lossy attacks such as JPEG compression, Gaussian noise, and other distortions. To enhance robustness against such distortions, a differentiable noise layer is introduced between the forward encryption and backward decryption to simulate real-world degradations during training. This layer includes JPEG (JPEGSS[Shin and Song 2017]), Gaussian noise, random crop, Gaussian blur, median blur, cropout, dropout and salt-and-pepper Noise. These differentiable operations allow the network to learn to resist common image corruption while maintaining end-to-end trainability.

\subsection{Loss Function}

To guarantee the effectiveness of the proposed framework, two different loss functions are introduced: the triplet recovery loss, which ensures lossless image reconstruction, and the cipher loss, which enforces encryption effect and visual randomness. 

\subsubsection{Triplet Recover Loss}

The aim of the backward decryption process is to accurately recover the plain image. To achieve this, Mean Squared Error (MSE) loss is applied to measure the discrepancy between images.To further enhance robustness, especially under incorrect key scenarios, a triplet loss is introduced to contrastly optimize correct and incorrect reconstructions. The full triplet recovery loss is defined as: 
\begin{equation}
 \begin{aligned}
        \mathcal{L}&_{TriRecover}(\theta)=
        \mathcal{L}_{TriMSE}(A,P,N) \\
         =&\mathcal{L}_{TriMSE}(I_P,I^{'}_P,I^{''}_P)  \\
         =&\text{max}\{0,MSE(I_P,I^{'}_P) - MSE(I_P,I^{''}_P) + 1.0\}
 \end{aligned}
\end{equation}
 where $\mathcal{L}_{TriMSE}$ denotes a triplet loss based on MSE distance metric, $A=I_P$ (anchor: the plain image), $P=I^{'}_P$ (positive: the correctly decrypted image), and  $N=I^{''}_P$ (negative: the wrong decrypted image using a wrong key). The triplet loss encourages the correct reconstruction to be closer to the plain image while push away the incorrect reconstruction, even in the absence of direct supervision for the latter. 

\subsubsection{Cipher Loss}

To ensure the security and visual unpredictability of the cipher image $I_C$ in the forward encryption process, a cipher loss is introduced, consisting of two components:
 \begin{equation}
    \mathcal{L}_{Cipher}(\theta)= \mathcal{L}_{uniform} + \mathcal{L}_{corr} 
 \end{equation}
where $\mathcal{L}_{uniform}$ represents uniform distribution loss which encourages the histogram distribution of $I_C$ to approximate a uniform distribution, thus reducing statistical redundancy: 
 \begin{equation}
    \mathcal{L}_{uniform} =D_{KL}(P(h_{I_C})||U) 
 \end{equation}
 where $D_{KL}$ denotes the Kullback-Leibler divergence, which is used to measure the similarity of two distributions, $P(h_{I_C})$ is the estimated histogram probability of $I_C$ over 256 gray levels, and $U=\frac{1}{256}$ denotes the target uniform distribution. Additionly, $\mathcal{L}_{corr}$ denotes pixel correlation loss which minimizes spatial correlation between adjacent pixels in $I_C$ to enhance visual randomness:
 \begin{equation}
         \mathcal{L}_{corr}
    =\vert \rho_{xy} \rvert 
    =\left|\frac{\sum_{i=1}^{N}(x_i-\bar{x})(y_i-\bar{y})}{\sqrt{\sum_{i=1}^{N}(x_i-\bar{x})^2 \times \sum_{i=1}^{N}(y_i-\bar{y})^2}}\right|
 \end{equation}
 where $x_i$ and $y_i$ denotes values of adjacent pixels in $I_C$, $N$ is the total number of the sampled pixel pairs, and $\bar{x}=1/N(\sum_{i=1}^N)x_i$,  $\bar{y}=1/N(\sum_{i=1}^N)y_i$ are their respective means.

\subsubsection{Total Loss}

The total loss function $\mathcal{L}_{Total}$ is a weighted sum of cipher loss $\mathcal{L}_{Cipher}$ and triplet recovery loss $\mathcal{L}_{TriRecover}$:
 \begin{equation}
    \mathcal{L}_{Total}(\theta)= \lambda_1\mathcal{L}_{Cipher} + \lambda_2\mathcal{L}_{TriRecover} 
 \end{equation}
where $\lambda_1$ and $\lambda_2$ are weight factors to balance encryption strength and reconstruction fidelity.

\section{Experimental Results}
\subsection{Experimental Settings}

\subsubsection{Datasets and settings}

The DIV2K\cite{DIV2K_2017} training dataset is used for training FlowCrypt, with training images resized to $112 \times 112$. For evaluation, we use three benchmark datasets: the DIV2K testing set (100 images), COCO \cite{lin2015microsoftcoco} (5,000 images), and ImageNet \cite{ImageNet_ILSVRC15} (50,000 images), with all test images resized to $256 \times 256$. The number of ACBs is set as 4 as it  achieves the trade-off between performance and model complexity. The parameters of $\lambda_1$ and $\lambda_2$ are fixed as 5.0 and 5.0, respectively. The entire training process takes about 4000 steps (approximately 20 hours) to stabilize. The framework is implemented by PyTorch\cite{Collobert_Pytorch_2011} and is trained using one single NVIDIA RTX 2080Ti GPU. Optimization is performed using Adam\cite{2015_Adam} with $\beta_1 = 0.9$ and $\beta_2 = 0.999$, and an initial learning rate of $1 \times 10^{-4.5}$.

\subsubsection{Benchmarks}

To verify the recovery quality and encryption security of proposed method, we compare it with several state-of-the-art (SOTA) image encryption methods, including the first INN-based end-to-end robust image encryption (ERIE) method \cite{INN_RIE_2025}, CNN-based method CNN-IE \cite{Erkan_2022_CNNRIE}, and a traditional chaotic encryption scheme, LTMM-CIEA \cite{HUA20212DLogisticTent}. For a fair comparison, all methods are evaluated on images resized to $256 \times 256$. 

\subsubsection{Evaluations}

To measure the quality of recovered images, we employ peak signal-to-noise ratio (PSNR), structural similarity index (SSIM) \cite{SSIM_2024}, root mean square error (RMSE), and mean absolute error (MAE). Higher PSNR and SSIM, along with lower RMSE and MAE, indicate better recovery quality. For encryption security, we analyze histogram and entropy. For key security evaluation, we adopt number of pixel change rate (NPCR) and uniform average intensity (UACI), with ideal values of 99.61\% and 33.46\%, respectively.

\subsection{Encryption and Decryption Effect}

\subsubsection{Qualitative Results}

\begin{figure}[!htbp]
\centering
\includegraphics[width=1.0\columnwidth]{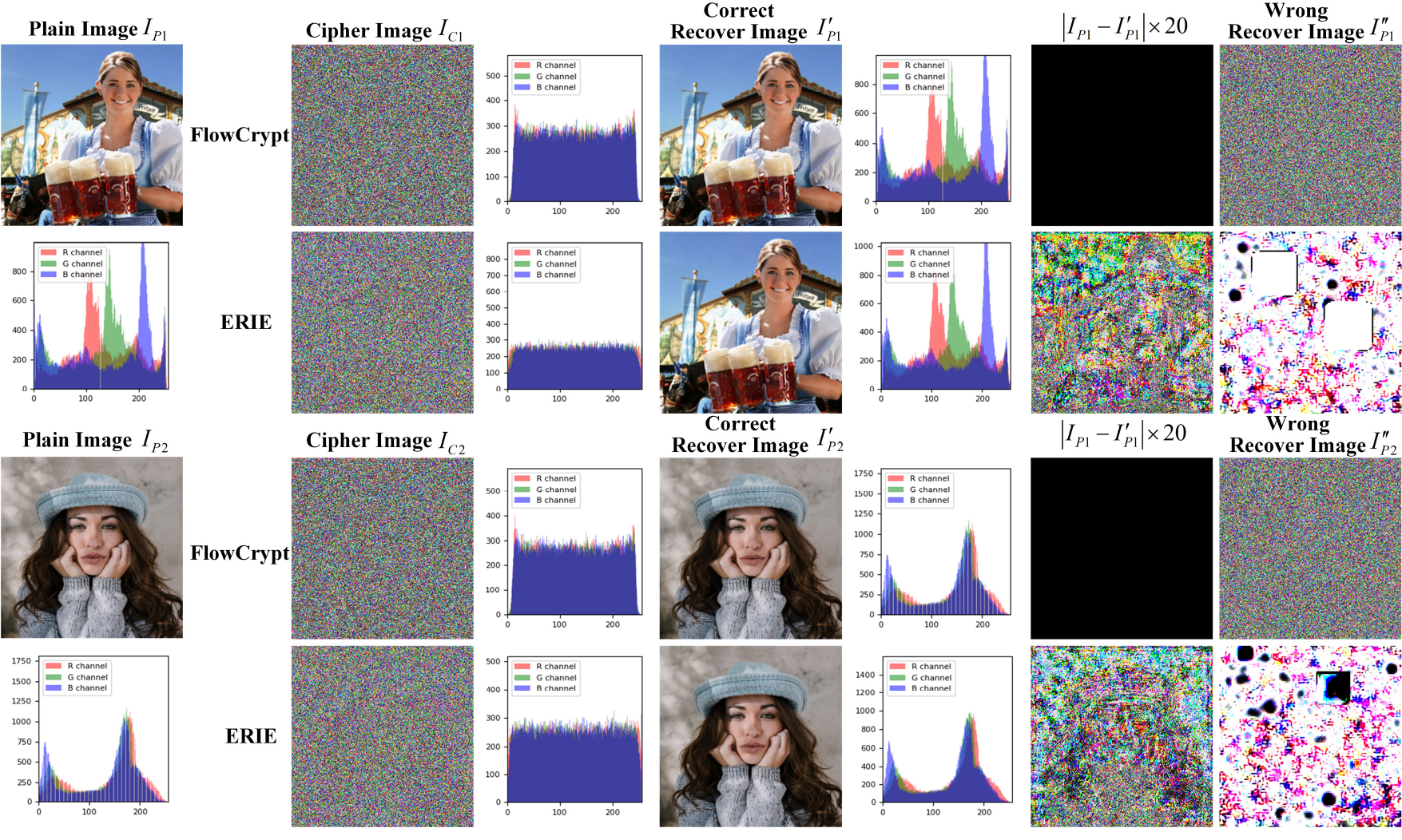} 
\caption{Visual comparison results of FlowCrypt and ERIE.}
\label{fig3}
\end{figure}

As aforementioned, the proposed FlowCrypt could achieve high-quality image recovery due to the inherent invertibility of the flow-based framework. As illustrated in Fig.~\ref{fig3}, we present visual comparisons results to present the superior encryption effect and near-lossless recovery. The encrypted cipher images appear as noise-like patterns with no discernible visual structure, and their corresponding histograms are uniformly distributed, effectively concealing semantic information. The correctly recovered images closely resemble the original plaint images, while the wrong recoveries display significantly distorted appearance. Furthermore, the residual images (magnified 20 times) between the correct recoveries and ground-truth plain images are obviously all black, indicating minimal reconstruction error. In contrast, residuals from ERIE exhibit noticeable artifacts, reflecting inferior recovery performance. These results confirm that FlowCrypt achieves both secure encryption effect and near-lossless reconstruction fidelity.

\begin{table*}[t]
\centering
\small
{
\begin{tabular}{c|cccc}
\toprule
\multirow{2}{*}{Method} & \multicolumn{4}{c}{DIV2K}  \\ 
\multirow{2}{*}{} & PSNR(dB)$\uparrow$ & SSIM$\uparrow$ & MAE$\downarrow$ & RMSE$\downarrow$ \\ 
\midrule
ERIE         & 51.97$\pm$7.76  & 0.9986$\pm$1.67e-03 & 1.69e-03$\pm$2.30e-04  & 2.56e-03$\pm$5.14e-04   \\  
FlowCrypt    & 115.78$\pm$0.09 & 0.9999$\pm$3.15e-06 & 1.08e-06$\pm$1.34e-08 & 1.62e-06$\pm$1.72e-08 \\  
\bottomrule
\multirow{2}{*}{Method} & \multicolumn{4}{c}{COCO}  \\ 
\multirow{2}{*}{} & PSNR(dB)$\uparrow$ & SSIM$\uparrow$ & MAE$\downarrow$ & RMSE$\downarrow$ \\ 
\midrule
ERIE         & 52.26$\pm$1.54 & 0.9986$\pm$1.02e-03  & 1.64e-03$\pm$2.08e-04   &   2.47e-03$\pm$4.84e-04 \\  
FlowCrypt    & 115.76$\pm$0.09 & 0.9999$\pm$2.36e-06    & 1.08e-06$\pm$1.29e-08 & 1.62e-06$\pm$1.76e-08 \\  
\bottomrule
\multirow{2}{*}{Method} &  \multicolumn{4}{c}{ImageNet} \\ 
\multirow{2}{*}{} & PSNR(dB)$\uparrow$ & SSIM$\uparrow$ & MAE$\downarrow$ & RMSE$\downarrow$ \\ 
\midrule
ERIE         & 51.86$\pm$0.92 & 0.9984$\pm$8.74e-04 & 1.64e-03$\pm$1.18e-04   & 2.56e-03$\pm$2.89e-04   \\  
FlowCrypt    & 115.76$\pm$0.09 & 0.9999$\pm$4.63e-06    & 1.08e-06$\pm$1.38e-08 & 1.62e-06$\pm$1.82e-08 \\  
\bottomrule
\end{tabular}
}
\caption{Benchmark comparisons of recovery performance.}
\label{table1}
\end{table*}

\subsubsection{Quantitative results}

Table \ref{table1} presents the quantitative comparison of FlowCrypt with ERIE. Each cell presents the mean value along with the corresponding standard deviation. As shown, FlowCrypt outperforms ERIE across all four metrics. Specifically, it achieves PSNR improvements of 63.81dB, 63.50dB, and 63.90dB over the same INN-based method ERIE on the DIV2K, COCO, and ImageNet datasets, respectively. Similar improvements are observed in SSIM, RMSE, and MAE, highlighting the superior reconstruction quality of FlowCrypt. These improvements stem from the reversible nature of the flow-based architecture, and preserves all information throughout the encryption-decryption process, enabling near-lossless recovery. Notably, although our model is trained solely on the DIV2K dataset, it generalizes well to COCO and ImageNet, demonstrating robust cross-domain performance, which is critical for practical applications such like cloud photo storage.

\begin{figure}[t]
\centering
\includegraphics[width=1.0\columnwidth]{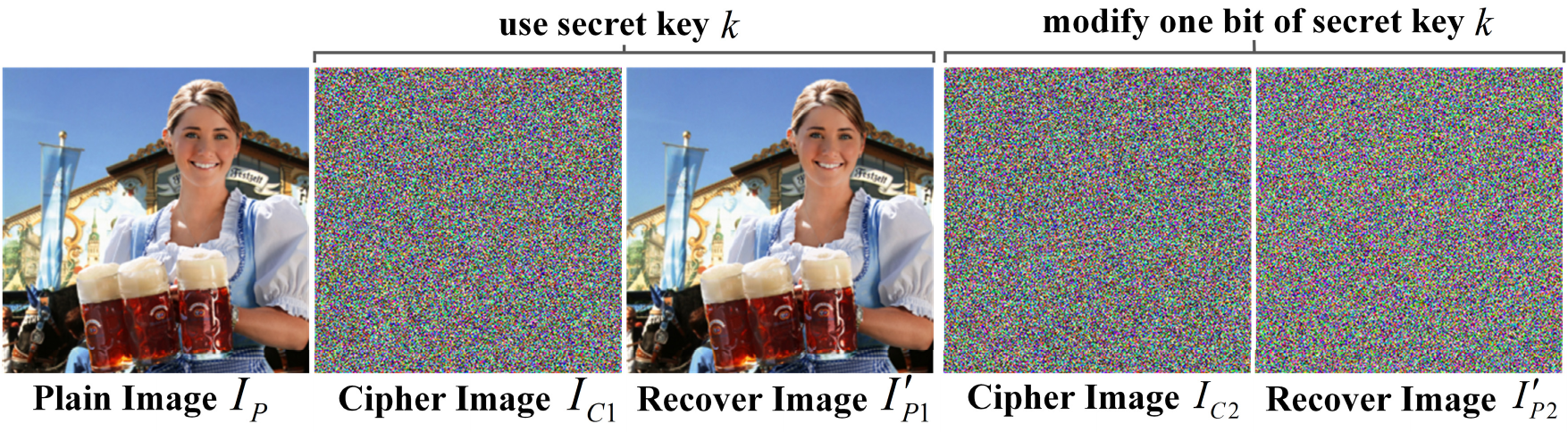} 
\caption{Key sensitivity analysis.}
\label{fig4}
\end{figure}

\begin{table}[]
\centering
\small
\setlength{\tabcolsep}{0.8mm}
{
\begin{tabular}{c|ccc}
\toprule
{Method} & NPCR (99.61\%) & UACI (33.46\%) & Entropy (8) \\ 
\midrule
CNN-IE       & 98.82 & 29.11 & 7.9989 (0.0011) \\  
ERIE         & 99.58 & 32.94 & 7.7558 (0.2442) \\  
FlowCrypt    & 99.38 & 32.52 & 7.9934 (0.0066) \\  
\bottomrule
{Method} &  Correlation (H)$\downarrow$ & Correlation (V)$\downarrow$ &  Correlation (D)$\downarrow$ \\ 
\midrule
CNN-IE       & 0.0011 & -0.0003 & 0.0034\\  
ERIE         & 0.0002 &  0.0038 & 0.0014\\  
FlowCrypt    & 0.0029 &  0.0001 & -0.0023 \\  
\bottomrule
\end{tabular}
}
\caption{Benchmark comparisons of security performance.}
\label{table2}
\end{table}

\begin{table*}[!htbp]
\centering
\small
\setlength{\tabcolsep}{0.5mm}
{
\begin{tabular}{c|*{5}{c}|*{5}{c}}
\toprule
\multirow{2}{*}{Method}  
& \multicolumn{5}{c|}{Cropout (\%)} 
& \multicolumn{5}{c}{Dropout (\%)} \\ 
\multirow{2}{*}{} & r=0.1 & 0.2 & 0.3 & 0.4 & 0.5 
& r=0.2 & 0.3 & 0.4 & 0.5 & 0.6 \\
\midrule
LTMM-CIEA    & 7.97/0.007  & 7.97/0.007  & 7.97/0.007  & 7.97/0.007  & 7.97/0.007  & 7.97/0.007  & 7.97/0.007  & 7.97/0.007  & 7.97/0.007  & 7.97/0.007  \\  
CNN-IE       & 27.94/0.916 & 22.12/0.745 & 18.55/0.574 & 16.09/0.440 & 14.11/0.331 & 15.10/0.380 & 13.34/0.283 & 12.09/0.216 & 11.12/0.164 & 10.33/0.123 \\  
ERIE         & 27.49/0.965 & 20.74/0.925 & 17.14/0.865 & 14.48/0.785 & 12.30/0.679 & 6.15/0.014  & 6.15/0.014  & 6.15/0.014  & 6.16/0.015  & 6.16/0.015  \\  
FlowCrypt    & 31.74/0.971 & 25.54/0.936 & 22.10/0.891 & 19.52/0.830 & 17.49/0.755 & 19.82/0.573 & 18.22/0.496 & 17.10/0.439 & 16.25/0.392 & 15.49/0.353 \\  
\bottomrule
\multirow{2}{*}{Method}  
& \multicolumn{5}{c|}{SP Noise (\%)} 
& \multicolumn{5}{c}{JPEG Compression (\%)} \\  
\multirow{2}{*}{} & r=0.01 & 0.02 & 0.03 & 0.04 & 0.05 
& Q=50 & 60 & 70 & 80 & 90 \\ 
\midrule
LTMM-CIEA    & 7.97/0.007  & 7.97/0.007  & 7.97/0.007  & 7.97/0.007  & 7.97/0.007  & 7.97/0.007  & 7.97/0.007  & 7.97/0.007  & 7.97/0.007  & 7.97/0.007  \\  
CNN-IE       & 28.12/0.886 & 25.10/0.804 & 23.33/0.742 & 22.08/0.692 & 21.12/0.651 & 11.53/0.104 & 11.92/0.116 & 12.28/0.128 & 12.61/0.140 & 13.06/0.157 \\  
ERIE         & 16.60/0.473 & 13.96/0.304 & 12.61/0.241 & 11.61/0.205 & 10.80/0.175 & 14.60/0.278 & 15.51/0.337 & 16.25/0.417 & 16.44/0.513 & 16.59/0.622 \\  
FlowCrypt    & 29.46/0.888 & 26.66/0.811 & 25.10/0.750 & 23.95/0.699 & 23.09/0.656 & 30.01/0.893 & 28.76/0.871 & 27.89/0.856 & 27.29/0.844 & 26.89/0.834 \\  
\bottomrule
\multirow{2}{*}{Method}  
& \multicolumn{3}{c|}{Gaussian Noise (\%)} 
& \multicolumn{3}{c|}{Gaussian Blur (\%)} 
& \multicolumn{4}{c}{Median Blur (\%)} \\  
\multirow{2}{*}{} & $\sigma=0.01$ & 0.03 & 0.05 
& \multicolumn{1}{|c}{$\sigma=0.5$} & \multicolumn{1}{c}{1} & 2 
& \multicolumn{1}{|c}{$w=3$} & 5 & 7 & 9  \\ 
\midrule
LTMM-CIEA   & 7.97/0.007  & 7.97/0.007  & 7.97/0.007  & \multicolumn{1}{|c}{7.97/0.007}  & \multicolumn{1}{c}{7.97/0.007}  & 7.97/0.007   & \multicolumn{1}{|c}{7.97/0.008}  & 7.97/0.007  & 7.97/0.008  & 7.97/0.008 \\  
CNN-IE      & 11.83/0.195 & 10.08/0.110 & 9.49/0.081  & \multicolumn{1}{|c}{9.77/0.093}  & \multicolumn{1}{c}{7.67/0.032}  & 7.87/0.011   & \multicolumn{1}{|c}{8.33/0.011}  & 8.14/0.007  & 8.12/0.007  & 8.12/0.007 \\  
ERIE        & 24.07/0.632 & 15.14/0.279 & 11.76/0.169 & \multicolumn{1}{|c}{23.82/0.721} & \multicolumn{1}{c}{13.52/0.152} & 8.7535/0.040 & \multicolumn{1}{|c}{15.11/0.206} & 9.06/0.044  & 7.74/0.024  & 7.39/0.019 \\  
FlowCrypt   & 25.93/0.796 & 21.72/0.644 & 19.92/0.564 & \multicolumn{1}{|c}{31.82/0.946} & \multicolumn{1}{c}{24.36/0.788} & 21.98/0.652  & \multicolumn{1}{|c}{24.90/0.797} & 23.51/0.695 & 22.44/0.626 & 21.61/0.579\\  
\bottomrule
\end{tabular}
}
\caption{Benchmark comparisons of recovery performance against common distortions, the metric is PSNR/SSIM in each cell.}
\label{table3}
\end{table*}

\subsection{Security and Complexity Analysis}

\subsubsection{Key Security}

We evaluate the security performance of FlowCrypt from three aspects: key space, key sensitivity and statistical characteristics of the cipher image. For key space, a larger key space enhances resistance to brute-force attacks. In FlowCrypt, the key consists of two components: (1) a random mask $M$ used to split the plain image, with a size of $2^{W*H}$; and (2) a secret map $\mathbf{K}$ that interacts with image features in the ACBs, with a size of $2^{W*H/2}$. Both the random mask $M$ and secret map $\mathbf{K}$ are deterministically derived from the user's password using PBKDF2. Therefore, the total key space is governed by the password entropy and PBKDF2 parameters, and can be expressed as $C^{L}$, where $C$ represents the size of the character set and $L$ is the length of the user's key. For instance, consider a commonly used password configuration: a 16-character key chosen from the ASCII character set, which includes 95 printable characters. In this case, the total key space is $95^{16} \approx 10^{31.64} \approx 2^{105.12}$. This expansive key space is sufficient to resist exhaustive brute-force attacks. 

In terms of key sensitivity, a secure encryption system should exhibit high sensitivity to key variations, which means that even a 1-bit change in the secret key should lead to significant differences in the recovered image. As shown in Fig.~\ref{fig4}, it is obvious that the recovery $I_{P2}^{'}$ fails completely when a 1-bit change in the secret key $k$, demonstrating the effectiveness of key sensitivity. To quantitatively assess key sensitivity, we use NPCR and UACI to test 1000 recovered images decrypted using 1-bit changed keys. As shown in Table~\ref{table2}, both NPCR and UACI values of FlowCrypt closely align with their ideal values, further confirming its strong key sensitivity and robustness against differential attacks. Additional results can be found in Appendix B.1.

As for statistical characteristics, we analyze the information entropy and pixel correlation of cipher images. Table~\ref{table2} lists the average results over 1000 cipher images. FlowCrypt achieves entropy values close to the theoretical ideal of 8, outperforming CNN-IE and ERIE, indicating uniformly distributed pixel values. Additionally, the average correlations in horizontal, vertical, and diagonal directions are all close to zero, suggesting minimal correlation between adjacent pixels and strong resistance to statistical analysis.

\begin{table}[!htbp]
\centering
\small
{
\begin{tabular}{c|cccc}
\toprule
{Method} & Params & Enc Time & Dec Time & Total Time \\  
\midrule
LTMM-CIEA    &   -   & 2.988 & 2.945 & 5.933 \\  
CNN-IE       &   -   & 3.951 & 3.983 & 7.934  \\  
ERIE         & 2.27M & 0.024 & 0.022 & 0.046  \\  
FlowCrypt    & 1.01M & 0.013 & 0.011 & 0.024 \\  
\bottomrule
\end{tabular}
}
\caption{Comparisons of computation complexity.}
\label{table4}
\end{table}

\subsubsection{Robustness Measurement}

During cloud transmission and storage, images may often subject to common degradations such as JPEG compression, Gaussian noise, Gaussian blur, median blur, cropout, dropout and salt-and-pepper (SP) Noise. To enhance robustness, FlowCrypt is trained using a differential noise layer inspired by MBRS \cite{MBRS_2021}. We evaluate the recovery quality under these seven distortions, as summarized in Table~\ref{table3}. Cropout refers to cropping a certain ratio of images and replacing it with zeros, while Dropout randomly drops pixel regions and replacing them with zeros. Across a range of crop ratios (0.1 to 0.5), FlowCrypt consistently achieves the best recovery quality, outperforming other methods in terms of robustness.

For noise distortions, FlowCrypt also demonstrates superior performance. Under SP noise with a ratio of 0.05, it maintains a recovery quality of 32.39dB. As for Gaussian noise, FlowCrypt outperforms other methods by 15.65dB. Similar advantages are observed under blurring distortions. For Gaussian blurring, FlowCrypt reaches a PSNR of 32.08dB. Even under strong median blur ($7 \times 7$ kernel), it maintains the best performance, surpassing other methods by 14.44dB. For JPEG compression, FlowCrypt is trained with the JPEGSS noise layer and evaluated across quality factors from 50 to 90. Consistently, it shows the highest robustness. These results highlight the effectiveness of the invertible network in interaction between encryption and decryption, contributing to robustness against distortions.

\subsubsection{Computation Complexity}

In cloud privacy scenarios, lightweight model size and computational efficiency are critical for deployment. We evaluate the average encryption and decryption time on 1000 color images, with results shown in Table~\ref{table4}. Compared to other methods, FlowCrypt achieves the smallest model size and the highest runtime efficiency, contributes to the lightweight deployment in edge devices such like mobile phone and vehicle equipments.

\begin{table}[!htbp]
\centering
{
\small
\setlength{\tabcolsep}{0.5mm}
\begin{tabular}{c|cc|cc}
\toprule
\multirow{2}{*}{Module} & \multicolumn{2}{c|}{Recovery Quality} &  \multicolumn{2}{c}{Encryption Effect} \\ 
\multirow{2}{*}{} & PSNR(dB)$\uparrow$ & SSIM$\uparrow$ & Entropy (8) & Correlation$\downarrow$\\ 
\midrule
Chessboard       & 80.07 & 0.9999 & 7.9552 & 0.2476 \\  
Random           & 99.99 & 1.0    & 7.9780 & 0.0968 \\  
Random + PBKDF   & 100.0 & 1.0    & 7.9934 & 0.0021 \\  
\bottomrule
\multirow{2}{*}{Module} & \multicolumn{2}{c|}{Recovery Quality} &  \multicolumn{2}{c}{Error Recovery Quality}\\ 
\multirow{2}{*}{} & PSNR(dB)$\uparrow$ & SSIM$\uparrow$ & PSNR(dB)$\downarrow$ & SSIM$\downarrow$ \\ 
\midrule
w/o Triplet Loss   & 100.0 & 1.0 & 11.09 & 0.6681 \\  
w/ Triplet Loss    & 100.0 & 1.0 & 9.19  & 0.5521 \\  
\bottomrule
\end{tabular}
}
\caption{Ablation study of split strategy and triplet loss.}
\label{table5}
\end{table}

\subsection{Ablation Study}

As shown in Table \ref{table5}, split strategy plays a critical role in improving the encryption effect of FlowCrypt. We compare three strategies: (1) chessborad split, which partitions the image in a regular grid; (2) random split using a fixed seed; and (3) random split with PBKDF, where a user password is transformed into a key. From the first to second row, entropy increases by 0.0228, and pixel correlation decreases by 0.1508, indicating the effectiveness of the random split strategy. The PBKDF-based split further improves histogram uniformity, and both entropy and pixel correlation approach ideal values, contributing to more secure key generation and cipher images. The triplet loss $L_{TriRecover}$ is designed to encourage accurate recovery while suppressing incorrect decryption outputs. As shown in Table \ref{table5}, $L_{TriRecover}$ further reduces the recovery quality of incorrect keys, thereby enhancing key sensitivity. More ablation studies can be found in Appendix B.2.

\section{Conclusion}
To address the limitations of existing deep learning-based encryption methods in recovery quality, robustness and practicality, this paper proposes FlowCrypt, a novel flow-based image encryption framework featuring near-lossless recovery, high security and lightweight design for cloud image storage. A key-conditioned random split strategy partitions the plain image in a stochastic and key-dependent manner, significantly improving encryption security. A parameter-shared, invertible encryption–decryption module ensures accurate recovery and high efficiency, without auxiliary data or information loss. Extensive experiments show that FlowCrypt achieves up to 100dB PSNR on three datasets, generates uniformly distributed cipher images, and maintains a compact model with only 1M parameters, making it suitable for mobile and edge-device deployments.

\bibliography{arxiv}

\end{document}